\documentclass[10pt,english,letterpaper]{article}
\usepackage[letterpaper]{geometry}
\usepackage{amsmath}
\usepackage{graphicx}
\usepackage{setspace}
\usepackage{amssymb}
\usepackage{esint}
\usepackage{indentfirst}
\usepackage{multicol}
\usepackage{graphicx,fancyhdr,amsmath,amsfonts,subfigure,psfrag}
\def\PACS#1{\par\noindent\strut\kern25pt{\small\rm PACS numbers: #1}\par}

\makeatletter
\newcommand{\etap}{\eta^{\prime}}

\newcommand{\pio}{\pi^0}
\newcommand{\kp}{K^+}
\newcommand{\kn}{K^-}
\newcommand{\kstp}{K^{*+}}
\newcommand{\kstm}{K^{*-}}
\newcommand{\ksto}{K^{*0}}

\newcommand{\ar}{\rightarrow}
\newcommand{\jpsi}{J/\psi}
\newcommand{\psip}{\psi(2S)}

\newcommand{\xeta}{X_{\eta}}
\newcommand{\yeta}{Y_{\eta}}

\newcommand{\xetap}{X_{\etap}}
\newcommand{\yetap}{Y_{\etap}}

\begin{document}
\title{Pseudoscalar mixing in $\jpsi$ and $\psip$
decay}

\author{
Dai-Hui Wei, Yong-Xu Yang\\
\small College of Physics and Technology, Guangxi Normal
University, Guilin 541004, China}

 \maketitle
\begin{abstract}
Based on the branching fractions of $\jpsi(\psip)\ar VP$ from
different collaborations, the pseudoscalar mixing is extensively
discussed with a well established phenomenological model. The
mixing angle is determined to be $-14^\circ$ by fitting to the new
world average if only quark content is considered. After taking
into account the gluonic content in $\eta$ and $\etap$
simultaneously, the investigation shows that $\eta$ favors only
consisting of light quarks, while the gluonic content of $\etap$
is $Z_{\etap}^2=0.30\pm0.24$.
\end{abstract}

\PACS{{13.25.Gv},{14.40.Ag}}
%
\section{Introduction}

\label{intro} As the ground pseudoscalar nonet, $\pi$, $K$, $\eta$
and $\etap$, in the constituent quark model, their masses and
widths are determined with high precision and the main decay modes
are also observed\cite{pdg2010} in addition to the forbidden and
rare decays. However there is one issue, pseudoscalar mixing,
remains not completely settled, which has been discussed for many
times with different transitions. The linear Gell-Mann-Okubo(GMO)
mass relation\cite{gmo} gives a mixing angle,
$\theta_P=-11^\circ$, which is hardly consistent with the value,
$\theta_P=-24.6^\circ$, obtained from the quadratic GMO mass
formula by replacing the meson masses by their squares. The full
set of $\jpsi$ decays into a vector and a pseudoscalar was
measured by MarkIII, and the phenomenological analysis of mixing
angle is determined to be
$\theta_P=(-19.2\pm1.4)^\circ$\cite{mark3}, which was confirmed by
DM2\cite{dm2}. Both of them got the conclusion that $\eta$ and
$\etap$ consist of light quarks, with no contribution from
gluonium or radial excitation states. After that an important work
was performed by Bramon and Scadron\cite{pvmeson2,pvmeson3},
taking into account $\omega$-$\phi$ mixing in the analysis for
$\jpsi\ar VP$, a weighted $\theta_P$ is calculated to be
$(-15.5\pm 1.3)^\circ$ based on many different transitions. For a
nice review based on the discussions before 2000, see\cite{review}
in which the reasonable range of $\eta$-$\etap$ mixing angle is
believed to  be $-20^\circ\sim -10^\circ$.

Recently the new experimental data on $\jpsi\ar VP$ and $\psip\ar
VP$ were reported by
BES\cite{besrhopi,besphip,besomegap,psipbesphip,psipbesomegap,psipbesrhopi,psipbeskstk},
BABAR\cite{babarrhopi,babaromegaeta,babarphieta,babarkstk} and
CLEO\cite{psipcleorhopi}. It is worth pointing out that some of
the new measurements are not well consistent with the previous
works. Take $\jpsi\ar\rho\pi$ for example, the branching fractions
measured by BES is $(2.10\pm0.12)\%$\cite{besrhopi}, subsequently
confirmed by BABAR\cite{babarrhopi}, which is larger than the
world average $(1.28\pm 0.10)\%$\cite{pdg1996}, about 64\%. This
significant change stimulates new interest in this
issue\cite{newvp,zhaoli,0802.3909,thomas}. The analysis in
Ref.\cite{0802.3909} indicates it is difficult to get reasonable
results with the updated branching fractions of $\jpsi\ar\rho\pi$,
however, the results in Ref.\cite{thomas} performed with the same
data and phenomenological model seems reasonable. This discrepancy
motivated us to reanalyze the full set of $\jpsi\ar VP$ data.
Actually it is difficult for us to compare the results obtained
with different sets of parameters in one time, in this paper we
would like to discuss this issue for different cases, e.g. fix
SU(3) breaking term $x$ to 0.64, 0.82 or 1.

\section{Notation}
\label{sec:1} The physical eigenstates $\eta$, $\etap$ are the
mixture of octet, singlet and gluonium. And they are defined as,
\begin{equation}
\begin{aligned}
|\eta>=X_\eta|N>+Y_\eta|S>+Z_\eta|G>,\\
|\etap>=X_\etap|N>+Y_\etap|S>+Z_\etap|G>.
\end{aligned}
\end{equation}
where, $N=\frac{1}{\sqrt{2}}(u\bar{u}+d\bar{d})$, $S=s\bar{s}$ and
$G$ for gluonium; $X_i$, $Y_i$ and $Z_i$ denote the magnitude of
non-strange, strange contents and gluonium in $\eta$ and $\etap$.

The above form can be written in terms of the three Euler angles,
with
\begin{equation}
\begin{aligned}
X_{\eta}=\cos\phi_p\cos\phi_{G1},\\
Y_{\eta}=-\sin\phi_p\cos\phi_{G1},\\
Z_{\eta}=-\sin\phi_{G1},\\
X_{\etap}=\cos\phi_P\cos\phi_{G2}-\sin\phi_P \sin\phi_{G2} \sin\phi_{G1},\\
Y_{\etap}=\sin\phi_P\cos\phi_{G2}+\cos\phi_P \sin\phi_{G2} \sin\phi_{G1},\\
Z_{\etap}=-\sin\phi_{G2}\cos\phi_{G1}.
\end{aligned}
\end{equation}

If we only consider the simplest case and neglect possible mixing
of the $\eta$ and $\etap$ with other pseudoscalar states,
$\eta$-$\etap$ mixing is characterized by a single mixing angle
$\theta_P$.
\begin{equation}
\begin{aligned}
|\eta>=\cos\theta_P|\eta_8>-\sin\theta_P|\eta_0>,\\
|\etap>=\sin\theta_P|\eta_8>+\cos\theta_P|\eta_0>.
\end{aligned}
\end{equation}
where $\eta$ and $\etap$ are the orthogonal mixture of the
respective singlet and octet iso-spin zero states. $\eta_{0}$ and
$\eta_{8}$ are SU(3) quark basis states which are denoted as
$\eta_{0}=\frac{1}{\sqrt{3}}|u\bar{u}+d\bar{d}+s\bar{s}>$ and
$\eta_{8}=\frac{1}{\sqrt{6}}|u\bar{u}+d\bar{d}-2s\bar{s}>$
respectively.

In terms of quark basis, the $\eta$ and $\etap$ include
non-strange and strange contents. In the flavor SU(3) quark model,
they are defined through quark-antiquark($q\bar{q}$) basis states
as,
\begin{equation}
\begin{aligned}
\xeta=\yetap=\sqrt{\frac{1}{3}}\cos\theta_P-\sqrt{\frac{2}{3}}
\sin\theta_p=\cos\phi_{P},\\
 \xetap=-\yeta=\sqrt{\frac{1}{3}}\sin\theta_P+
\sqrt{\frac{2}{3}}\cos\theta_P=\sin\phi_{P}.
\end{aligned}
\end{equation}
where, $\theta_P=\phi_P-54.7^\circ$.

\section{Phenomenological model}

$\jpsi$ and $\psip$ have the similar decay mechanism and are
suppressed by Okubo-Zweig-Iizuka (OZI) rule. Both of them decay
into a vector and pseudoscalar meson via three gluon annihilation
and electromagnetic decays. Therefore, in this paper, the
phenomenological model for $\jpsi\ar VP$ in Ref.\cite{haber} is
simply applied in $\psip$ decays to discuss the $\eta$-$\etap$
mixing and other physics.

A first-order parameterization of the amplitudes appears in
Ref.\cite{haber} and is described there in detail. The amplitude,
which has contributions from both the three gluon annihilation
 and electromagnetic processes
can be expressed in terms of an SU(3) symmetric single-OZI(SOZI)
amplitude $g$, an electromagnetic amplitude $e$ (the coupling
strength $e$ has a relative phase $\theta_e$ to the strength $g$
because these are produced from different origins)
 and the nonet-symmetry-breaking
double-OZI(DOZI) amplitude $r$, relative to $g$. SU(3) violation
has been accounted for by a pure octet SU(3) breaking term. The
SU(3) breaking term in strong interaction and electromagnetic
process are expressed by (1-$s$) and $x$, respectively. A factor
(1-$s$) for every strange quark contributing to $g$ and a factor
for $x$ for a strange quark contributing to $e$. The factor
$s_v$($s_p$) is for the strange vector(pseudoscalar) contributing
to $r$.

In spite of these simplified assumptions this phenological model
contains a rather large number of parameters ($g$, $e$, $r$, $s$,
$s_p$, $s_v$, $x$, $\theta_P$ and $\theta_e$). This $x$ can be
well determined via $V\ar P\gamma$ and $P\ar V\gamma$ data, we
reanalyzed it using the phenomenological model in Ref.\cite{vpr}
and the branching fractions of $V\ar P\gamma$ and $P\ar V\gamma$
in Ref.\cite{pdg2010}, $x$ is determined to be $0.82\pm0.05$ and
$\theta_V=(3.2\pm 0.9)^\circ$ and $\theta_P=(-12.9\pm0.5)^\circ$
which are in good agreement in those in Ref.\cite{newvp}. To
further simplify it again, $s_p$ is ignored in this paper and
$s_v$ is discussed below with two assumptions($s_v=0$ and
$s_v=s$).

\begin{table*}[htpb]
\caption{Branching fractions of $\jpsi\ar VP$($\times 10^{-4}$)}
\begin{center}
\begin{tabular}{c c c c c c}
\hline Decay Modes            &  MarkIII& DM2
  & BES  & BABAR & PDG2010 \\
\hline $\rho\pi$    & $142\pm 1\pm 9$   &$132\pm20$
              & $210\pm12\pm 20.1$   &$218\pm19$& $169\pm15$\\
$\rho\eta$   &$1.93\pm0.13\pm0.29$ &$1.94\pm0.17\pm0.29$ & & &$1.93\pm0.23$ \\
$\rho\etap$  &$1.14\pm0.14\pm0.16$&$0.83\pm0.30\pm0.12$  & & &$1.05\pm0.18$\\
$\phi\pio$   & $<0.068$               &  & $<0.064$               &  &$<0.064$\\
$\phi\eta$   & $6.61\pm0.45\pm0.78$ &$6.4\pm0.4\pm1.1$ & $8.98\pm0.24\pm0.89$ &$14\pm6\pm1$ &$7.5\pm0.8$ \\
$\phi\etap$  &$3.08\pm0.34\pm0.36$ &$4.1\pm0.3\pm0.8$ &$5.46\pm0.31\pm0.56$ & &$4.0\pm0.7$ \\
$\omega\pio$  &$4.82\pm0.19\pm0.64$&$3.6\pm0.28\pm0.54$ &$5.38\pm0.12\pm0.65$& &$4.5\pm0.5$  \\
$\omega\eta$  &$17.1\pm0.8\pm2.0$ &$14.3\pm1.0\pm2.1$  &$23.52\pm2.73$ &$14.4\pm4.0\pm1.4$ &$17.4\pm2.0$\\
$\omega\etap$ &$1.66\pm0.17\pm0.19$&$1.8^{+1.0}_{-0.8}\pm0.3$ &$2.26\pm0.43$&    &$1.82\pm0.21$\\
$\kstm\kp+c.c.$ &$52.6\pm1.3\pm5.3$& $45.7\pm1.7\pm7.0$ &   & $52\pm4\pm1$ &$51.2\pm3.0$\\
$\ksto \bar{K^{o}}+c.c.$& $43.3\pm 1.2\pm 4.5$&$39.6\pm1.5\pm6.0$ &  &$48\pm5\pm1$ &$43.9\pm3.1$ \\
\hline
\end{tabular}
\end{center}
\label{branch}
\end{table*}

\section{Results}
The experimental data sets shown in Table~\ref{branch} are
analyzed with the least squares method to determine the coupling
strengths and mixing angle. To clarify the results obtained from
different data set, we divided it into several subsections to
investigate the pseudoscalar mixing.
\subsection{Analysis of $\jpsi\ar VP$ from MarkIII and DM2 }
We start to perform the fit to experimental data with the simplest
case, the $\omega$-$\phi$ mixing and gluon content are ignored.
Actually the treatment on SU(3)-breaking parameter $x$ and the
second order corrections $s_v$ in Ref.\cite{mark3} and
Ref.\cite{dm2} are different. The $x$ is set to 1 and the
correction terms $s_v=0$, is ignored in MarkIII analysis, while
$x$ is fixed to 0.64 and the correction terms $s_v=s$ is included
in DM2 analysis. To clearly compare the difference between them,
all the possible combinations are considered to perform the fit.

A fit to the data without considering SU(3) breaking as well as in
MarkIII analysis yields $\theta_P=(-13.95\pm2.39)^\circ$ with
$\chi^2/d.o.f=9.0/4$, which is obviously inconsistent with the
value $(-19.2\pm 1.4)^\circ$\cite{mark3}. After tuning the
parameter, we also get a reasonable results which are the same as
those in
Ref.\cite{mark3},\\
$g=1.10\pm 0.03$, $s=0.12\pm 0.03$, $e=0.122\pm0.005$,
$\theta_e=1.25\pm 0.12$, $\theta_p=(-19.34\pm 1.40)^\circ$,
$r=-0.15\pm0.09$.\\
But the goodness of fit, $\chi^2/d.o.f=10.1/4$, seems slightly
worse. Compared with the results listed in the first column of
Table~\ref{mark3fit}, $s$ and $r$ also change significantly. The
results of the fits performed with $x=0.64$ and $x=0.82$ are also
given in Table~\ref{mark3fit}. Apart from the mixing angle, the
values of other parameters are also consistent with the previous
fit.

If $s_v$ is replaced with $s$ and $x$ is fixed to 0.64, the fit
gives $\theta_P=(-18.59\pm1.40)^\circ$ with $\chi^2/d.o.f=9.0/4$
which is in good agreement with DM2's result
$\theta_P=(-19.1\pm1.4)^\circ$. Meanwhile we also checked the fits
with $x=1$ and $x=0.82$ and the results are listed in
Table~\ref{mark3fit}. Compared with the results without
considering the contribution of $s_v$, the results change
significantly, in particular for $s$, $r$ and $\theta_p$. This is
reasonable because the two phenomenological models are slightly
different. The fit to DM2 data is also performed to check the
discrepancy discussed above. In the DM2's analysis, the common
error of the branching fractions is removed, so the fitting error
here is larger than those in Ref.\cite{dm2}. Here it is clear that
the reasonable results can also be obtained
$\theta_P=(-14.84\pm4.35)^\circ$, with $\chi^2/d.o.f=1.9/4$ in the
case of $s_v=0$ and $x=1$.

Based on the above results, we can get the conclusion that $s_v$
plays an important role in the fit to extract the mixing angle.
The mixing angle in DM2 analysis is consistent with that in
MarkIII because  the latter is not from the best fit.
\subsection{Analysis of $\jpsi\ar VP$ from BES, BABAR and PDG2010}
Until now the pseudoscalar mixing is investigated with the well
established models and the data measured about 20 years ago. The
new measurements reported by BES, BABAR and the new world average
of 2010 are listed in Table.\ref{branch}. Each branching fraction
is regarded as one constraint in the fit to BES and BABAR data.
The amplitude of $\jpsi\ar\rho\eta$ and $\jpsi\ar\rho\etap$ is
removed from the fit because no new measurements are available.
The results of the fits with $s_v=0$ yields the mixing angle
$\theta_p\sim -17^\circ$, which is still consistent with the above
results within one standard deviation. This value is also in
agreement with the previous work in
Ref.\cite{newvp,zhaoli,thomas}. The fit with $s_v=s$ is performed,
but the quality of fit is very poor.

The further check is performed using the world average of
2010\cite{pdg2010}, and the results are shown in
Table~\ref{pdg2008fit}. As we expected, the results are fine for
the fit with $s_v=0$ and the mixing angle $\theta_p$ favors $\sim
14^\circ$. The goodness of the fit with $s_v=s$ is still worse
because of the weight of new measurements in the world average.

\subsection{Analysis of $\psip\ar VP$}
We now turn to examine the full set of $\psip\ar VP$ to get the
pseudoscalar mixing using the same phenomenological model. At
present the measurements of $\psip\ar VP$ mainly come from BES and
CLEO's reports which are shown in Table~\ref{psipbranch}. We have
omitted the known upper limit for the $\psip\ar\phi\pi$ and
$\psip\ar\omega\eta$ branching fractions in our analysis because
they are the upper limits at 90\% confidence level rather than
branching fractions. As previously stated, we just consider the
mixing angle between $\eta$ and $\etap$ and assume the mixing of
$\omega$ and $\phi$ is ideal. The results listed in
Table~\ref{psipfit} indicate that both of the above two slightly
different models are reasonable and the $\theta_p$ favors
$\sim-12^\circ$ with large uncertainty. Without considering the
branching fraction of $\psip\ar\rho\pi$, the fit was also
performed in Ref.\cite{thomas}, the mixing angle is calculated to
be $-10^{+7}_{-8}$ which is in agreement with our result. But the
branching fraction of $\psip\ar\kstp\kn$, $(8.5\pm 4.0)\times
10^{-5}$, applied in the analysis is not correct. Therefore the
values of parameters listed in Table~\ref{psipfit} are
inconsistent with those in Ref.\cite{thomas}.

\begin{table*}[htpb]
\caption{Results of fit to MarkIII data}
\begin{tabular}{c c c c c c c}
\hline Parameter   &$s_v=0$,x=1 & $s_v=0$,x=0.64 & $s_v=0$,x=0.82
            &$s_v=s$,x=1 & $s_v=s$,x=0.64 & $s_v=s$,x=0.82\\
\hline
 g &        $1.30\pm0.04$ &$1.31\pm0.04$ &$1.30\pm0.04$
   &        $1.12\pm0.04$ &$1.11\pm0.04$ &$1.11\pm0.04$\\
 s &        $0.27\pm0.02$ &$0.28\pm0.02$ &$0.27\pm0.02$
   &        $0.13\pm0.03$ &$0.13\pm0.02$ &$0.13\pm0.03$\\
 e &        $0.124\pm0.005$&$0.123\pm0.05$&$0.124\pm0.05$
   &        $0.123\pm0.005$&$0.123\pm0.005$&$0.123\pm0.005$\\
 $\theta_e$&$1.21\pm0.12$  &$1.29\pm0.12$&$1.27\pm0.12$
   &        $1.27\pm0.12$  &$1.30\pm0.12$  &$1.29\pm0.12$\\
 r &        $-0.37\pm0.01$ &$-0.37\pm0.01$&$-0.37\pm0.01$
   &        $-0.16\pm0.01$ &$-0.15\pm0.01$ &$-0.15\pm0.01$\\
$\theta_P$ &$-13.95\pm2.39$&$-13.17\pm2.40$&$-13.49\pm2.38$
   &        $-18.29\pm1.43$& $-18.59\pm1.40$&$-18.47\pm1.41$\\
 $\chi^2/d.o.f$ &9.0/4      & 7.9/4  & 8.3/4
               &8.1/4      & 9.0/4  & 8.6/4\\
\hline
\end{tabular}
\label{mark3fit}
\end{table*}

\begin{table*}[htpb]
\caption{Results of fit PDG2010 data}
\begin{tabular}{c c c c c c c}
\hline Parameter   &$s_v=0$,x=1 & $s_v=0$,x=0.64 & $s_v=0$,x=0.82
            &$s_v=s$,x=1 & $s_v=s$,x=0.64 & $s_v=s$,x=0.82\\
\hline
 g &        $1.35\pm0.04$ &$1.36\pm0.04$ &$1.36\pm0.04$
   &        $1.15\pm0.04$ &$1.14\pm0.04$ &$1.14\pm0.04$\\
 s &        $0.30\pm0.02$ &$0.30\pm0.03$ &$0.30\pm0.02$
   &        $0.15\pm0.03$ &$0.14\pm0.03$ &$0.14\pm0.03$\\
 e &        $0.120\pm0.005$&$0.121\pm0.04$&$0.121\pm0.04$
   &        $0.119\pm0.005$&$0.119\pm0.005$&$0.119\pm0.005$\\
 $\theta_e$&$1.31\pm0.12$  &$1.36\pm0.12$&$1.34\pm0.12$
   &        $1.35\pm0.12$  &$1.38\pm0.12$  &$1.36\pm0.12$\\
 r &        $-0.37\pm0.01$ &$-0.37\pm0.01$&$-0.37\pm0.01$
   &        $-0.16\pm0.01$ &$-0.15\pm0.01$ &$-0.15\pm0.01$\\
$\theta_P$ &$-14.27\pm2.44$&$-13.90\pm2.35$&$-14.04\pm2.37$
   &        $-17.66\pm1.81$& $-17.96\pm1.77$&$-17.84\pm1.78$\\
$\chi^2/d.o.f$ &3.1/4      & 3.5/4  & 3.3/4
               &16.5/4      & 18.1/4  & 17.4/4\\
\hline
\end{tabular}
\label{pdg2008fit}
\end{table*}

\begin{table*}[htpb]
\caption{Branching fractions of $\psip\ar VP$($\times 10^{-5}$) }
\begin{center}
\begin{tabular}{c c c c}
\hline Decay modes            & BES & CLEO & PDG2010 \\
\hline
$\rho\pi$    & $5.1\pm0.7\pm 1.1$   &$2.4\pm0.8\pm0.2$& $3.2\pm1.2$\\
$\rho\eta$   & $1.78^{+0.67}_{-0.62}\pm0.17$ &$3.0^{+1.1}_{-0.9}\pm0.2$&$2.2\pm0.6$ \\
$\rho\etap$  &$1.87^{+1.64}_{-1.11}\pm0.33$ &            &$1.9^{+1.7}_{-1.2}$ \\
$\phi\pio$   & $<0.4$               &$<0.7$  &$<0.4$ \\
$\phi\eta$   & $3.3\pm1.1\pm0.5$ &$2.0^{+1.5}_{-1.1}\pm0.4$ &$2.8^{+1.0}_{-0.8}$ \\
$\phi\etap$  &$3.1\pm1.4\pm0.7$ &                  &$3.1\pm1.6$\\
$\omega\pio$
&$1.87^{+0.68}_{-0.62}\pm0.28$&$2.5^{+1.2}_{-1.0}\pm0.2$
&$2.1\pm0.6$ \\
$\omega\eta$  &  $<3.1$     &$<1.1$ & $<1.1$ \\
$\omega\etap$ &$3.2^{+2.4}_{-2.0}\pm0.7$&    &$3.2^{+2.5}_{-2.1}$  \\
$\kstm\kp+c.c.$ &$2.9^{+1.3}_{-1.7}\pm0.4$
&$1.3^{+1.0}_{-0.7}\pm0.3$&$1.7^{+0.8}_{-0.7}$ \\
$\ksto \bar{K^{o}}+c.c.$& $13.3^{+2.4}_{-2.8}\pm1.7$
&$9.2^{+2.7}_{-2.2}\pm0.9$
&$10.9\pm2.0$\\
\hline
\end{tabular}
\end{center}
\label{psipbranch}
\end{table*}

\begin{table*}[htpb]
\caption{Results of fit to PDG2010 data of $\psip\ar VP$}
\setlength{\tabcolsep}{3pt}
\begin{tabular}{c c c c c c c}
\hline Parameter   &$s_v=0$,x=1 & $s_v=0$,x=0.64 & $s_v=0$,x=0.82
            &$s_v=s$,x=1 & $s_v=s$,x=0.64 & $s_v=s$,x=0.82\\
\hline
 g &        $0.64\pm0.11$ &$0.65\pm0.10$ &$0.65\pm0.10$&$0.64\pm0.11$
&$0.65\pm0.04$
&$0.65\pm0.04$\\
 s &        $0.003\pm0.18$ &$-0.01\pm0.19$ &$-0.05\pm0.18$
   &        $0.02\pm0.18$ &$-0.10\pm0.19$ &$-0.10\pm0.20$\\
 e &        $0.23\pm0.02$&$0.23\pm0.02$&$0.23\pm0.02$
   &        $0.23\pm0.02$&$0.23\pm0.02$&$0.23\pm0.02$\\
 $\theta_e$&$2.73\pm0.62$  &$2.81\pm0.60$&$2.79\pm0.63$
   &        $2.75\pm0.64$  &$2.83\pm0.64$  &$2.83\pm0.62$\\
 r &        $0.18\pm0.28$ &$0.17\pm0.27$&$0.16\pm0.28$
   &        $0.14\pm0.28$ &$0.14\pm0.29$ &$0.14\pm0.31$\\
$\theta_P$ &$-12.07\pm10.42$&$-11.94\pm10.48$&$-11.99\pm10.46$
   &        $-11.80\pm10.63$& $-12.19\pm11.59$&$-12.19\pm12.18$\\
$\chi^2/d.o.f$ &4.4/3      & 4.5/3  & 4.4/3
               &4.4/3      & 4.5/3  & 4.5/3\\
\hline
\end{tabular}
\label{psipfit}
\end{table*}

\subsection{$\omega$-$\phi$ mixing}
In the above analysis, the $\omega$-$\phi$ mixing is ignored to
simplify the model. This fit in the case of $s_v=0$ and $x=0.82$
is an attempt to account for the $\omega$-$\phi$ mixing. If
$\omega$-$\phi$ mixing angle is left as a free parameter, the fit
to the world average of 2010 leads to a minimum $\chi^2=3.3$ for
three degrees of
freedom,\\
$g=1.36\pm 0.04$, $s=0.30\pm 0.03$, $e=0.121\pm0.005$,
$\theta_e=1.33\pm 0.12$, $\theta_p=(-14.06\pm2.37)^\circ$,
$r=-0.37\pm0.02$, $\theta_V=(0.09\pm4.13)^\circ$.\\
If we assumed $s_v=s$, then the fit
with $\chi^2/d.o.f$ of 17.4/3 gives,\\
$g=1.14\pm 0.05$, $s=0.14\pm 0.04$, $e=0.119\pm0.005$,
$\theta_e=1.36\pm 0.12$, $\theta_p=(-17.78\pm2.70)^\circ$,
$r=-0.15\pm0.01$, $\theta_V=(0.11\pm 3.78)^\circ$.

$\theta_V$ is very close to zero and the uncertainty is very large
compared with other parameters. This means that there is not a
significant constraint on it. Among $\jpsi \ar VP$ decays, the
amplitude of $\jpsi\ar\phi\pio$ is directly related to the
$\omega$-$\phi$ mixing, but it is still not observed yet. No
observation of $\jpsi\ar\phi\pio$ shows that the contribution of
$\omega$-$\phi$ is small. On the other hand, the values of other
parameters are almost the same as those listed in
Table~\ref{pdg2008fit} without considering $\omega$-$\phi$ mixing.
Therefore it is reasonable that $\omega$-$\phi$ mixing is assumed
to be ideal and could be ignored in the above analysis. Further
check is done by fixing the $\omega-\phi$ mixing angle to
$3.2^\circ$ obtained from $V\ar
\gamma P$ and $P\ar \gamma V$  process. The fit  with $\chi^2/d.o.f=3.3/4$ gives,\\
$g=1.36\pm 0.04$, $s=0.30\pm 0.02$, $e=0.121\pm0.004$,
$\theta_e=1.33\pm 0.13$, $\theta_p=(-14.05\pm2.36)^\circ$,
$r=-0.37\pm0.01$, these values are also in good agreement with
those in the hypothesis of the ideal $\omega$-$\phi$ mixing.

\subsection{Gluon content in $\eta$ and $\etap$}
At present $\eta$ is believed to be well-understood as an SU(3)
flavor octet with a small quarkonium singlet admixture, and not
much room for a  significant gluonium
admixture\cite{newvp,thomas}. Therefore the analyses\cite{thomas}
are usually performed to determine the gluonic content in $\etap$
with the assumption of no gluonic content in $\eta$. After taking
into account the gluonic content in $\eta$ and $\etap$
simultaneously, we present the fit with the above two slightly
different models. In the first case, $s_v$ is assumed to be zero
and the fit
to the world average in 2010 yields,\\
$g=1.32\pm 0.06$, $s=0.27\pm 0.04$, $e=0.126\pm0.007$,
$\theta_e=1.34\pm 0.12$, $\theta_p=(-10.21\pm 4.48)^\circ$,
$r=-0.45\pm0.08$, $\phi_{g1}=0.04\pm0.05$,
$\phi_{g2}=0.53\pm0.24$,
$r^{\prime}=-0.77\pm 0.46$,\\
with $\chi^2/d.o.f=1.56/1$.

The second fit is performed under the hypothesis of $s_v=s$, and
the results with
$\chi^2/d.o.f =3.5/1$ are listed as follows,\\
$g=1.28\pm 0.06$, $s=0.24\pm 0.03$, $e=0.128\pm0.007$,
$\theta_e=1.35\pm 0.11$, $\theta_p=(-9.17\pm 4.67)^\circ$,
$r=-0.67\pm0.08$, $\phi_{g1}=0.11\pm0.04$,
$\phi_{g2}=0.50\pm0.22$, $r^{\prime}=-0.85\pm 0.56$.

The goodness of the second fit is still worse than the first fit.
Based on the results of the first fit, the magnitudes of gluon
components in $\eta$ and $\etap$ are calculated to be
$Z^{2}_{\eta}=0.002\pm0.002$ and $Z^{2}_{\etap}=0.30\pm0.24$,
respectively. The small gluonic contribution in $\eta$ shows there
is not much room for gluonium admixture, which is consistent with
the results presented in Ref.\cite{zhaoli}. It seems that 30\% of
$\etap$ component could be attributed to gluonium, but further
investigation with more precisely data needed to be done due to
the large uncertainty.

\section{Summary and outlook}

A wide set of data on $\jpsi\ar VP$ and $\psip\ar VP$ decays are
analyzed in terms of a rather general phenomenological model in an
attempt to determine the magnitudes of components in $\eta$ and
$\etap$. The data include the branching fractions of $\jpsi\ar VP$
which were measured nearly 20 years ago and the recent
measurements by BES and BABAR. The measurements of MarkIII and DM2
are reanalyzed, we found that the results obtained from the two
different phenomenological models are inconsistent. The fit to the
new measurements by BES and BABAR indicates that the assumption of
$s=s_v$ is not a good approximation in accordance with the
goodness of fit. And the mixing angle is determined to be
$-14^\circ$, which is in good agreement with previous works.

The content of $\eta$ and $\etap$ is also examined in this paper.
After considering the gluonium content in the model, the fit to
data of the world average in 2010 yields
$Z^{2}_{\eta}=0.002\pm0.002$ and $Z^{2}_{\etap}=0.30\pm0.24$,
which are the contribution of
 gluonium
content in $\eta$ and $\etap$ respectively. Although the
possibility of gluonic content can not be excluded, it is a
reasonable description for $\eta$ in terms of pure $q\bar{q}$
meson, and no much room for a  significant gluonium admixture. The
magnitude of gluonium contamination in $\etap$ shows that $\etap$
has room for gluonium admixture, but the large uncertainty
prevents us from definitely saying that gluonium content is
present or not.

As we previously stated that the latest results from BES and BABAR
are not consistent with those previous works. The branching
fractions shown in Table~\ref{branch} still have large error,
including statistical and systematic errors. The main reason is
that $\jpsi$ and $\psip$ samples are not enough and the
performance of detector need to be improved. A modern detector,
BESIII\cite{besiii}, has been built to meet the above
requirements. Up to now, about $2.3\times 10^{8}$ $\jpsi$ and
$1.2\times 10^{8}$ $\psip$ events have been accumulated at BESIII,
which provide a unique chance to study the $\eta$-$\etap$ mixing
and further improve these measurements with much higher
sensitivities.





\clearpage
\end{document}